\documentclass[useams,useamsmath,usenatbib]{mnras}
\usepackage{amsmath}
\usepackage{amssymb}
\usepackage{setspace}
\usepackage{pdfpages}

\def\eg{\,{\rm e.g.}\,}

\def\ccite{\cite}
\def\ccitep{\citep}

\def\bbf{}
\begin{document}
\twocolumn
\title[Stellar-mass Evolution]{GAMA/G10-COSMOS/3D-HST: Evolution of the galaxy stellar mass function over 12.5Gyrs}
\author[Wright et al.]
{A.H.~Wright$^{1}$\thanks{e-mail:awright@astro.uni-bonn.de},
S.P.~Driver$^{2,3}$,
A.S.G.~Robotham$^{2}$ \\
$^{1}$Argelander--Institut f\"ur Astronomy, Universit\"at Bonn, Auf dem H\"ugel 72, D-53121 Bonn, Germany \\
$^{2}$ICRAR\thanks{International Centre for Radio Astronomy Research}, The
University of Western Australia, 35 Stirling Highway, Crawley, WA 6009, Australia\\
$^{3}$SUPA, School of Physics\& Astronomy, University of St Andrews, North Haugh, St Andrews, KY16 9SS, UK \\
}
\date{Received XXXX; Accepted XXXX}
\pubyear{2018} \volume{000}
\pagerange{\pageref{firstpage}--\pageref{lastpage}}
\maketitle
\label{firstpage}
\begin{abstract}
  Using a combined {and consistently analysed } GAMA, G10-COSMOS, and 3D-HST dataset 
  we explore the evolution of the galaxy
  stellar-mass function over lookback times $t_{\rm L} \in
  \left[0.2,12.5\right] {\rm h^{-1}_{70} Gyr}$.  We use a series of volume
  limited samples to fit Schechter functions in bins of $\sim\!$constant
  lookback time and explore the evolution of the best-fit parameters in both
  single and two-component cases.  In all cases, we employ a fitting procedure
  that is robust to the effects of Eddington bias and sample variance.
  Surprisingly, when fitting a two-component Schechter function, we find
  essentially no evidence of temporal evolution in $M^\star$, the two $\alpha$
  slope parameters, or the normalisation of the low-mass component.  Instead,
  our fits suggest that the various shape parameters have been exceptionally
  stable over cosmic time, as has the normalisation of the low-mass component,
  and that the evolution of the stellar-mass function is well described by a
  simple build up of the high-mass component over time.  When fitting a single
  component Schechter function, there is an observed evolution in both
  $M^\star$ and $\alpha$, however this is interpreted as being an artefact.
  Finally, we find that the evolution of the stellar-mass function, and the
  observed stellar mass density, can be well described by a simple model of
  constant growth in the high-mass source density over the last $11 {\rm
  h^{-1}_{70} Gyr}$. 
  \vspace{0.5in}
\end{abstract}
\section{Introduction}\label{sec: intro} 
Understanding the redshift evolution of galactic properties is a fundamental method
for understanding the growth and evolution of structure over cosmic times. These
studies typically explore integrated galaxy parameters such as stellar mass density
\ccitep[$\rho_\star$, e.g.][]{Madau2014}, galaxy population morphological parameters
such as the early-type fraction \ccitep[e.g.][]{Davidzon2017}, individual galaxy
evolution parameters such as star formation rates \ccitep{Driver2017}, environmental
parameters such as the galaxy two-point correlation function 
\ccitep[e.g.][]{Croom2005,Zheng2007} and merger rate \ccitep[e.g. ][]{Bridge2010}, and
formation parameters such as the galaxy halo mass function \ccitep[e.g.][]{Moster2010}.  
All of these parameters encode complex physics about the formation and growth of galaxies 
over time.

The integrated galaxy stellar mass density is of particular interest as it can be
directly compared to the integrated cosmic star formation history \ccitep[see
][ for an extensive review of such studies]{Madau2014}. Over the last decade
deep near-IR surveys such as the Great Observatories Origins Deep Survey 
\ccitep[GOODS;][]{Giavalisco2004}, the MUlti-wavelength Survey by Yale-Chile 
\ccitep[MUSYC;][]{Gawiser2006}, the Cosmic Assembly Near-infrared Deep Extragalactic Legacy
Survey \ccitep[CANDELS;][]{Grogin2011,Koekemoer2011}, the Cosmic Evolution Survey
\ccitep[COSMOS;][]{Scoville2007} and the FourStar Galaxy Evolution Survey 
\ccitep[ZFOURGE;][]{Tomczak2014}, have made studying the properties of high redshift galaxies
increasingly accessible to the astronomical community. Meanwhile, large
multi-wavelength surveys of the low-redshift universe such as the Galaxy And Mass
Assembly \ccitep{Driver2011,Driver2016}, the 2dF Galaxy Redshift Survey
\ccitep{Cole2001}, and Sloan Digital Sky Survey \ccitep{Bell2003} have allowed us to
explore galaxy properties with high statistical accuracy out to redshifts of
$\sim\!0.5$. By combining datasets from these two classes of surveys, we are able to
create combined samples that allow us to explore, with high number statistics
throughout, the evolution of galactic parameters over a vast redshift range.

In this work, we explore the evolution of the shape of the galaxy stellar mass
function (GSMF) using a combination of surveys. In Section \ref{sec: data} we
describe the dataset used here. In Section \ref{sec: method} we describe the
parameterisation of the GSMF and the fitting methods employed. In Section 
\ref{sec: results} we present the results of our analysis, with a discussion of the
implications of our fits in Section \ref{sec: discussion}, and in Section 
\ref{sec: conclusions} we provide some concluding remarks.

Throughout this work we use a concordance cosmological model of $\Omega_{\rm M}=0.3$,
$\Omega_{\Gamma}=0.7$, ${\rm H_0}=70\,{\rm km s^{-1} Mpc^{-1}}$ and ${\rm h_{70}} =
{\rm H_0}/70 \, {\rm km s^{-1} Mpc^{-1}}$.  All masses are derived/quoted using a
time-invariant \ccite{Chabrier2003} IMF, \ccite{Bruzual2003} population synthesis
models, and a \ccite{Charlot2000} dust model. {\bbf All magnitudes are presented in the AB system}. 
\section{Data}\label{sec: data} 
In this paper we use the combined GAMA \ccitep{Driver2009,Driver2011,Driver2017},
G10-COSMOS \ccitep{Davies2015,Andrews2017}, and 3D-HST
\ccitep{Brammer2012,Momcheva2016,Skelton2014} dataset described in
\ccite{Driver2017}. The dataset combines these three highly complementary surveys
which comprehensively sample the low-, mid-, and high-redshift universe respectively.
Importantly, each dataset has been analysed in a consistent way, making this dataset
somewhat distinct from other large compilations of data presented and analysed in the
literature \ccitep[e.g.][]{Rodriguez2017}. The dataset primarily comprises of
spectral energy distributions (SEDs) for all galaxies in each of these surveys, and
with this dataset \ccite{Driver2017} were able to compile a statistically
representative sample of galaxies for 19 consecutive bins of redshift spanning the
range $z\in\left[0.02,5.00\right]$, which equates to lookback times of $t_{\rm trav}
\in \left[0.28,12.31\right]\,{\rm h_{70}^{-1}\,Gyr}$.

A full description of the dataset, including a discussion of systematic effects such
as redshift and magnitude completeness, is available in \ccite{Driver2017}. {
Nonetheless here we provide a brief overview of the dataset, including summarising 
the methods of our photometric and SED analyses, before describing the methods of 
analysis used in this work. 

The GAMA and G10-COSMOS data described in \ccite{Driver2017} both utilise
photometry measured using the {\sc lambdar} software \ccitep{Wright2016} in 21
and 22 filters respectively.  {\sc lambdar} is a bespoke photometric program
which was developed specifically to address the challenge of measuring
consistent matched aperture photometry on images that are neither seeing nor
pixel matched, while performing robust deblending, sky removal, and uncertainty
estimation. This is achieved within {\sc lambdar} by deblending seeing
convolved input apertures, generated independently for each input-image pixel
grid, with neighbouring galaxies and/or contaminants in an iterative manner.
Uncertainties are calculated incorporating shot noise, robust sky estimation
using per-galaxy local annuli, and estimation of correlated noise using
per-galaxy local blanks. Full description of the processes employed by {\sc
lambdar} can be found in \ccite{Wright2016}, and the program's application to
GAMA and G10-COSMOS can be found in \ccite{Wright2016} and
\ccite{Andrews2017} respectively.  Photometry in the 3D-HST fields are
downloaded from the 3D-HST website (\url{http://3dhst.research.yale.edu}),
rather than being calculated directly from the imaging using {\sc lambdar}.
This represents the only part of our analysis from photometry to final mass
function estimation that involves possibly inconsistent measurement methods. 
Our three datasets are generated from each survey using simple optical flux-limits, 
prior to the application of mass limits to generate volume-complete samples. 
For GAMA, G10-COSMOS and 3D-HST these optical selection limits are: $r<19.8$mag, 
$i\le25$mag and F$814$W$\le26.0$mag. 

Prior to mass estimation we perform an additional cleaning of AGN contaminated sources within 
the G10-COSMOS and 3D-HST datasets using the formalism described by \cite{Donley2012}, 
as these sources can lead to biases in our mass estimates. We also reject radio loud sources 
using the prescription of \cite{Seymour2008} in the G10-COSMOS sample. Finally, completeness of 
the variable-depth 3D-HST dataset is verified by comparison to deep HST number counts from the literature, 
and is found to be highly complete (within expected cosmic variance) to our adopted magnitude limit of F$160$W$=26.0$mag.
Each of these selections is described in detail in \cite{Driver2017}. 

Each of these photometric datasets is then fit using the SED modeling program
{\sc magphys} \ccitep{daCunha2008} using spectroscopic redshifts (for: all
GAMA, some G10-COSMOS, some 3D-HST), GRISM redshifts (for some 3D-HST), and/or
photometric redshifts (for: most G10-COSMOS, most 3D-HST). {\sc magphys} 
performs an energy balance of observed stellar-origin light with that emitted
from warm- and cold-dust, in order to recover the unobscured galaxy stellar
mass from observed fluxes. For our fits, we implement the standard {\sc
magphys} template library for each dataset, and then perform an additional run
of our 3D-HST dataset using the high-redshift {\sc magphys} template set.  The
best fitting template, as determined by the template with the minimum $\chi_{\rm r}^2$,
for each 3D-HST source is then selected. In practice, 97 percent of the 3D-HST sources 
are optimally fit by a standard {\sc magphys} template. 

{\sc magphys} outputs per-galaxy
posterior probability distributions for each model parameter, which we then use
for parameter inference. Specifically, we} use the median of the {\sc magphys}
posterior samples for all parameter inferences, and use the $16^{\rm th}$ to
$84^{\rm th}$ percentile range of the posterior samples to define each
parameter's uncertainty. {Full description of the application of {\sc
magphys} to this combined dataset is given in \ccite{Driver2017}.}

SED measurements of the galaxies in each dataset provide us with per-galaxy estimates
of the stellar mass, star formation history, and dust mass, although in this work we
focus solely on the estimated stellar masses.  Given the different survey areas and
depths, we can utilise these data to generate the stellar mass function over a range
of redshifts: the wide area of the GAMA survey provides good sampling of, in
particular, the high mass end of the mass function at low redshift, and the deep but
narrow G10-COSMOS and 3D-HST surveys provide significant information of the low mass
end of the GSMF at low redshift. These deep studies then transition to providing
information about the higher mass end of the mass function at high redshift. In this
work, as in \ccite{Driver2017}, we use only volume-complete samples of the full
dataset at each redshift interval, thus significantly reducing the possible number of
systematic biases that may affect our analysis\footnote{We note that recently codes
have been developed that significantly simplify the task of performing robust
analyses of non-volume limited datasets, such as the {\tt dftools} package presented
by \ccite{Obreschkow2018}. Nonetheless, we opt to continue with only volume limited
datasets, and leave this more sophisticated analysis for a future work.}. {Mass
completeness limits in each of the datasets have been calculated assuming that
the galaxy stellar mass functions exhibit no considerable down-turn over the
masses probed in this work. Using this {\bbf assumption-driven approach},
per-dataset per-bin completeness limits used in this work were estimated by
\ccite{Driver2017} by truncating each dataset to masses exclusively above where
a downturn in the mass function is observed.}{\bbf This assumption-driven approach 
is less rigorous than other methods of mass completeness estimation in the literature
\citep[see, e.g.,][]{Marchesini2009,Muzzin2013,Ilbert2013,Tomczak2014,Skelton2014}, 
however is unlikely to bias our analysis over the mass ranges 
we explore. This method of estimating mass 
completeness limits is likely to cause us to over-estimate the mass down to which 
our samples are complete, as, if a down-turn does exist, this will be interpreted as 
incompleteness and our analysis truncated prior to this downturn. Additionally, 
this method is vindicated in the first 13 bins of our analysis by the overlap
between the various datasets, and is further bolstered at low-redshift where our GAMA 
mass limits are in reasonable agreement with more rigorously determined limits for the same 
sample \citep{Baldry2018,Wright2017}. Nonetheless, the assumption is most susceptible to 
error in the highest redshift and lowest mass sections of our analysis; precisely the 
parts where we have no additional corroborating data. As such, we must recognise the 
possibility that the analysis in these areas is susceptible to error.}

\begin{figure*}
\centering
\includegraphics[width=\textwidth]{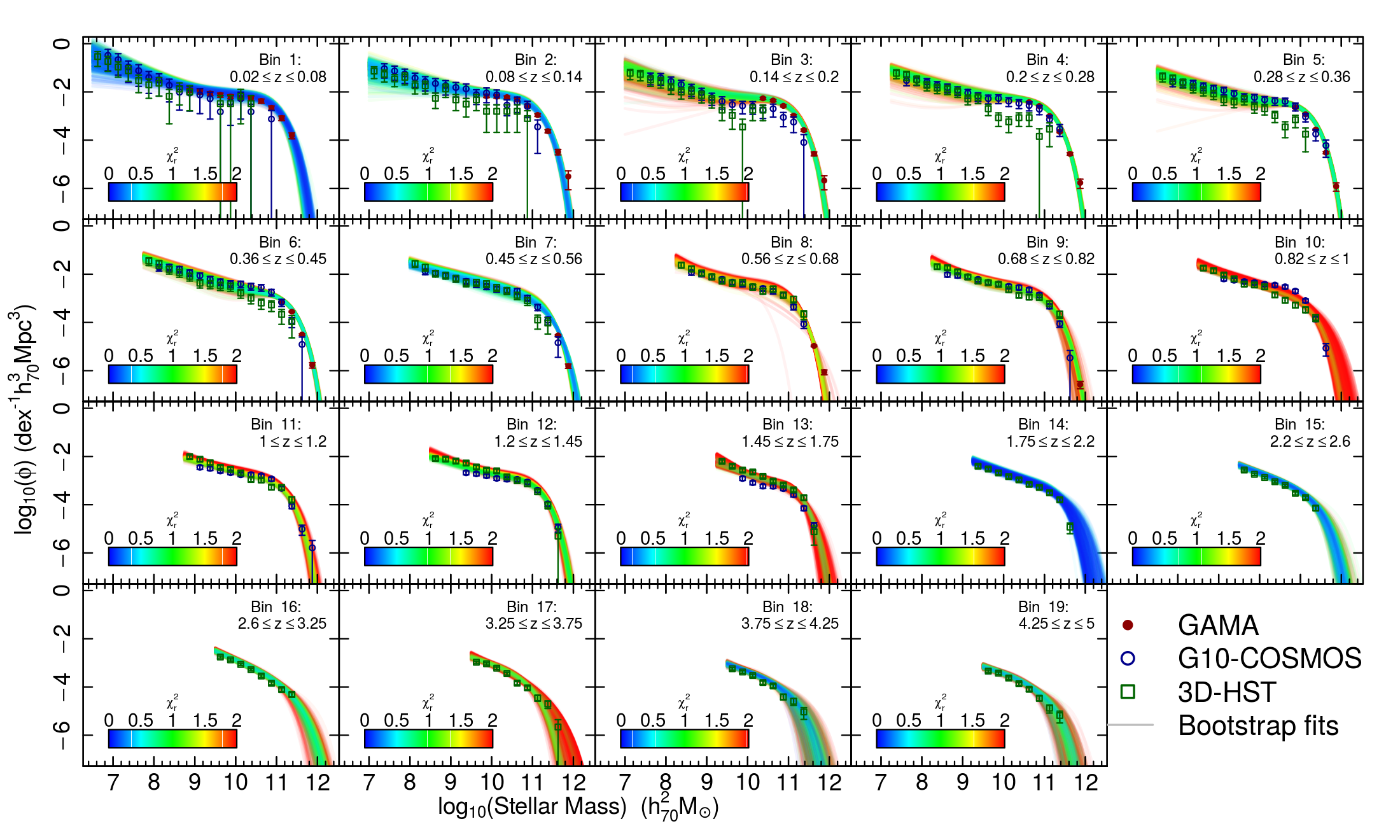}
\caption{The two-component Schechter function fits to our combined GAMA (red),
  G10-COSMOS (blue), and 3D-HST (green) dataset. Each panel shows a redshift bin
  (limits are annotated) with the fitted data, per survey, and the results from
  the bootstrapped fitting procedure (lines; coloured by fit reduced $\chi^2$).
  {Uncertainties on the datasets are determined by the cosmic variance uncertainty
  of each survey within the relevant redshift interval, and the shot noise per mass bin.
  Redshift} bins 3 and 4 show
  considerable drops in number density at high masses in G10-COSMOS and 3D-HST.
  However, even in these cases, the optimisation is able to borrow strength from
  the complementary GAMA dataset and converge on appropriate fits.
  }\label{fig: data bins}
\end{figure*}

Figure \ref{fig: data bins} shows our combined dataset for each of the redshift bins, with
shot and sample-variance uncertainties {indicated as the per-bin error bars}. 
Each panel shows the observed
number density for each of the surveys, binned in stellar mass, along with the
individual {monte-carlo+bootstrap} fits coloured by reduced $\chi^2$ (more details on this are
given in Section \ref{sec: method}). These panels highlight the value of this
combined dataset: even in regions where one or two of our datasets are lacking in
completeness (due to, for example, poor sampling due to a limited survey area; see
the 3D-HST data in redshift bin 4), we have sufficient complementarity between the
three surveys that there is no difficulty in estimating the completely free
two-component \ccite{Schechter1976} function.

\section{Fitting the galaxy stellar mass function}\label{sec: method} 
Fitting the \ccite{Schechter1976} function to our combined dataset requires careful
consideration of each individual dataset's sample variance uncertainty and selection
bias. To account for these, we invoke a fitting method that allows each sample to be
fit with its own mass limit and independent perturbation of the normalisation
(according to the expected sample variance). 

We also wish to incorporate our ignorance of precise stellar masses, and the expected
Eddington bias of our samples, into the fitting procedure as well.  To do this we
invoke a {combined monte-carlo+bootstrap simulation} method, applied
in bins of redshift, using the following steps at
each realisation. Starting from the raw data, we select those data within
the redshift bin, and perturb every source's stellar mass according to our {\sc
magphys} fit uncertainty.  {We then make a bootstrap realisation of this perturbed sample, 
which results in our fit data for this realisation. Next} we bin the three surveys by stellar mass
individually, discarding bins below each survey's mass limit, and divide the number
counts by the volume probed, per survey, over this redshift interval. We then perturb
each set of binned data by the expected sample variance uncertainty as reported by
\ccite{Driver2017}. These binned data points are then fit with a Schechter function
using the quasi-Newton optimisation algorithm of \ccite{Byrd1995}, which allows
box-constraint of optimisation parameters. We select these box constraints using
previous results from the literature. For all fits we provide box constraints of
$M^\star \in \left[10,11.5\right]$ and $\phi^\star \in \left[0,1\right]$. For our
single-component fits we also constrain $\alpha \in
\left[-2,1.5\right]$, and in the two-component case the individual $\alpha$'s are
required to be $\alpha_1 \in \left[-1.1,1.5\right]$ and $\alpha_2 \in
\left[-2,-0.9\right]$, thereby ensuring that the two components do not flip places
during optimisation, and discouraging fits with degenerate components.   
The resulting fit parameters are stored, with uncertainties
derived from the optimisation hessian matrix, and the next {realisation} is begun. We
perform $1001$ of these {combined monte-carlo+bootstrap} realisations per redshift bin\footnote{
 Testing with $10001$ realisations in our $3^{\rm rd}$ bin produced no change 
in parameter inferences}. For our final fit parameters
and uncertainties we take the 1/$\chi^2$-weighted median of all converged fit
parameters, and use the similarly weighted $16^{\rm th}$ to $84^{\rm th}$ percentile
range for our parameter uncertainties.  By using an optimisation procedure such as
this, we are able to simultaneously fit all of our three surveys' data. This allows
for better optimisation than would be possible by fitting each sample independently,
as the three highly complementary surveys provide constraints of different parts of
the Schechter function.

We explore the observed evolution of Schechter function parameters for both a single
and two-component Schechter function, and in both cases have all parameters free
(within the box-limits specified above).  In each of our optimisations, we maximise
the likelihood of the data with respect to a convolved version of the Schechter
function (with $\sigma_{\rm conv}=0.1$dex) to account for Eddington bias in our
samples \ccitep{Driver2017}. 

The two-component Schechter function is visually a much better description of the
data at low-redshift, {\bbf as can be seen by the clear plateau in the mass
functions}, and has been adopted almost unanimously as the appropriate
descriptor of the GSMF there 
\ccitep[see, e.g.,][]{Baldry2008,Baldry2012,Davidzon2017,Wright2017}. 
Conversely, at high-redshift the GSMF is {\bbf frequently argued} to be well
described by a single component Schechter function, which is often used for fitting 
mass functions there \ccitep[see, e.g.,][]{Song2016,Grazian2015,Davidzon2017}. 
{\bbf In our analysis, rather than assume a particular Schechter function formalism for 
different redshift bins, we opt 
to fit both single and two-component functions to each of our redshift bins, and 
provide fit parameters and goodness of fit statistics for each. By presenting both 
datasets in this way, we aim to explore how the mass function evolves under both 
assumptions without possibly uncertain restrictions.}  

At low redshift, the combination
of highly-complete GAMA data and the deep G10-COSMOS and 3D-HST data allows us to
simultaneously constrain the exponential cutoff of the GSMF (principally
parameterised by the $M^\star$ parameter), and the slope parameter(s) $\alpha_{i}$.
However at high redshift our constraint on the slope parameter is less robust, as the
data only extend to $\sim\! 1$dex below $M^\star$. In these higher-redshift bins, one
might expect that the two-component fits would become extremely noisy, as the
optimisation has far too much freedom given the data; {\bbf another reason to perform 
optimisations using both functional forms. We note, however, that
our choice of limiting values on the normalisation parameter allows the optimiser to
explore single component solutions even in the two-component optimisation. All individual 
fit parameters and reduced $\chi^2$ values are presented in Appendix \ref{ap: results}. }

In addition to fitting our two different Schechter forms, we also make some further
assumptions that allow us to better constrain the Schechter function form at each
redshift interval.  If we assume that redshift evolution of each parameter should
a-priori be a smooth function, then we can fit the redshift evolution of each
parameter with a simple function and use this to generate a less-noisy estimate of
how the Schechter function evolves over cosmic time. Therefore, after establishing
our best-fit Schechter parameters in each redshift bin, we also fit a quadratic
function to the redshift evolution of each parameter, and show the Schechter function
evolution using these best-fit functions.  This regression fit to the individual
parameters is not largely different to other (iterative) regression procedures
invoked in the literature \ccitep[see, e.g.,][]{Drory2008,Leja2015}.  We shown our
regressed fits (and uncertainties) alongside our individual optimisations, and the
fit parameters are also presented in Appendix \ref{ap: results}.


\section{Results}\label{sec: results} 
\begin{figure*}
  \centering
  \includegraphics[width=\textwidth]{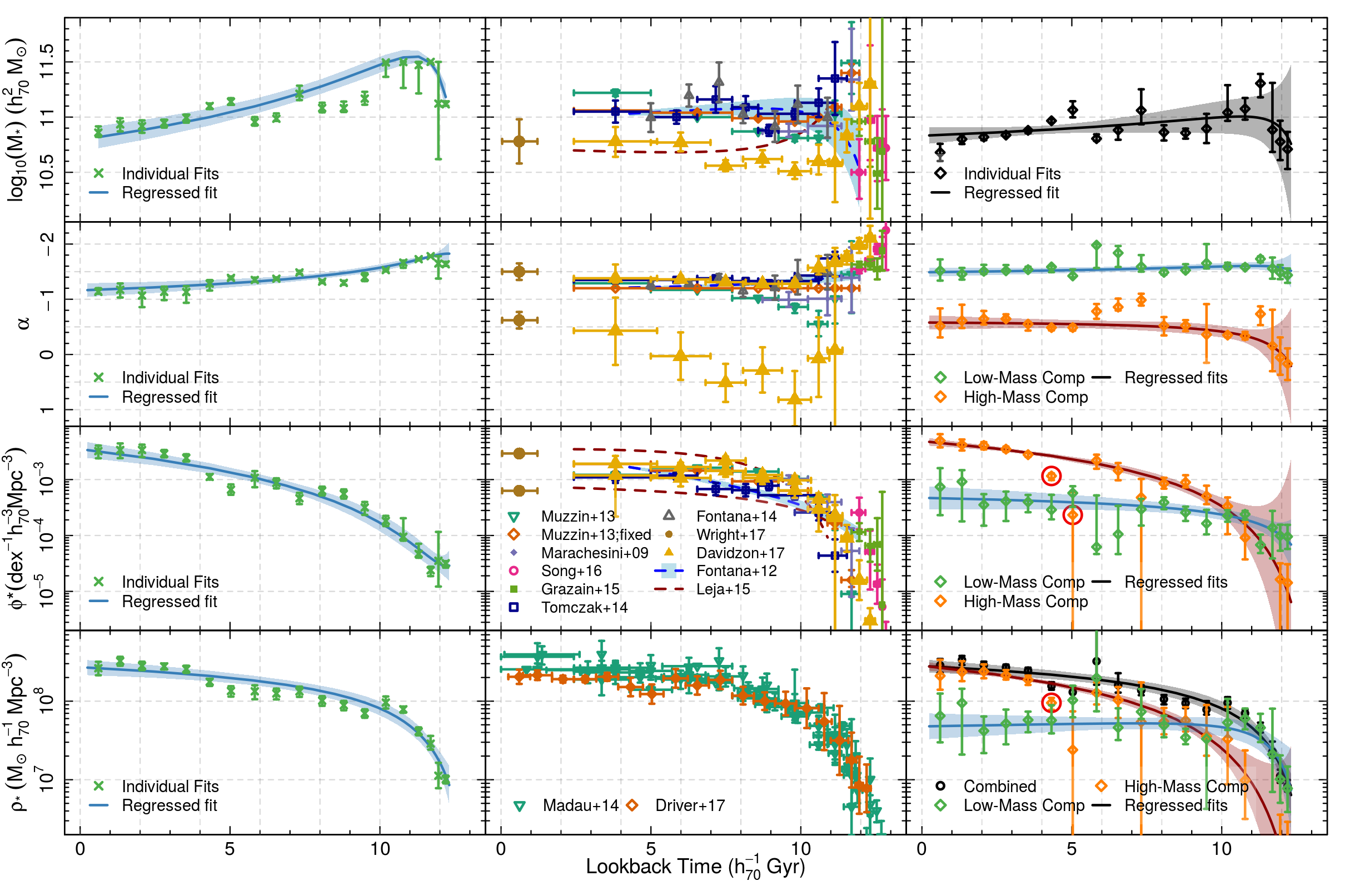}
  \caption{Evolution of the Schechter function parameters, and of the integrated
  stellar mass density, as a function of lookback time. {\em Center column:} Literature
  compendium of the evolution of each parameter over time. Relevant sources to the
  Schechter parameter evolution panels are annotated in the $\phi^\star$ panel. In
  $\rho_\star$, we show only the compendium of \protect \ccite{Madau2014} and the
  previous results using this dataset from \protect \ccite{Driver2017}. {\em Left
  column:} Results from our single component Schechter function fits to the dataset.
  Individual optimisations per bin are shown as points, with uncertainties showing the
  $\pm1\sigma$ confidence region determined from the individual bootstrap fits
  weighted by their $\chi_{\rm r}^2$. Our regressions are shown as lines with a shaded
  uncertainty region. {\em Right column:} Results from our two-component Schechter
  function fits to the dataset. Points and lines here are the same as in the center
  column, except now there are multiple components being shown in the $\phi^\star$
  and $\alpha$ panels. Points circled in red are clipped prior to the estimation of
  the regression fit, as they are in high-tension with the other data. All data
  points here are provided as supplementary data with this paper. Regression fits are
  given in Appendix \protect \ref{ap: results}. 
  }\label{fig: results}
\end{figure*}
Figure \ref{fig: results} shows the evolution of our Schechter function parameters
$M^\star$, $\phi^\star$, and $\alpha_i$, as well as the evolution of the integrated
stellar mass density ({ which is derived using the analytical integration of the Schechter function 
fits over all masses}).  In the figure, we
show a compilation of literature values for each parameter (center), as well as our
single (left) and two-component (right) fits, separated to aid clarity.  The
individual data are shown with uncertainties, as described in Section \ref{sec:
method}. The regression fits are shown with the uncertainty regions also shown as
shaded regions around the best fit line. We can see that, in all cases, the fits are
best constrained in the low-to-mid redshift bins, and that the fit uncertainties
increase significantly beyond lookback times of $\sim\! 11\,h_{70}^{-1}\,{\rm Gyr}$. 

Looking first at our two-component fits, we see a surprising lack of evolution in all
the shape parameters within our fits.  Our regressed fit in $M^\star$ is consistent
with being flat, although a pragmatic interpretation would likely be that there has
been a very slight decrease in the value of $M^\star$ over the last $\sim\!
11.5\,h_{70}^{-1}\,{\rm Gyr}$. Our fit also exhibits a downturn at higher redshifts,
however the constraint here is sufficiently weak that interpreting this as a real
feature is difficult. 

The single component fit shows a somewhat higher $M^\star$ than the two-component fit
at essentially all times, indicating a bias that can be induced when fitting single
component Schechter functions (even high redshift) to data that should likely be fit
with more components. The fits also move to significantly higher values of $M^\star$
at early times, causing the regression to behave somewhat poorly. We note that this
trend is also evident in the literature; studies that have invoked a two-component
Schechter function \ccitep[\eg][]{Leja2015,Wright2017,Davidzon2017} show
systematically lower values of $M^\star$ than those that fit only a single component
\ccitep[\eg][]{Fontana2006, Tomczak2014}. This trend is exacerbated by the
degeneracy between $M^\star$ and $\alpha$, which gets stronger as $\alpha$ approaches 
a value of -2 (as it does in the high-redshift single component fits). For these
reasons alone, we believe that there is a clear motivation to describe the shape of
the Schechter function with two components 
\ccitep[at least; see][]{Moffett2016,Kelvin2014}, even out to high redshifts.

The two-component slope parameters $\alpha_i$ also shows little evidence of
evolution. The low-mass component in particular is impressively stable over cosmic
time, showing only a minor flattening over the last $11 \,h_{70}^{-1}\,{\rm Gyr}$.
Conversely, our single component fits (and the single component regressed model) show
an appreciable evolution, and one that shows appreciable steepening of the mass
function slope (particularly at high-redshift). We argue that this observed evolution
is an artefact. At low-redshift the mass function is poorly described by a single
component fit, and the slope parameter is flattened by the plateau of the mass
function. Interestingly, at high-redshift, our single-component fits prefer  the same
increase in slope and $M^\star$ as is often seen in the literature, while our
two-component fits show no such effect.  This result is seen particularly well in the
recent work of \ccite{Davidzon2017}, who see their Schechter function slope and
$M^\star$ grow significantly steeper and more massive, respectively, in their highest
3 redshift bins where they transition from a two- to single-component fit. Meanwhile
our observed fits are in agreement with other studies that push to the estimation of
the GSMF to high redshift \ccitep{Song2016,Grazian2015}.  Finally, our high-mass
component shows a slight evolution to a steeper slope over the last $11
\,h_{70}^{-1}\,{\rm Gyr}$, however is also reasonably consistent with no evolution. 

The value of $\phi^\star$ shows the strongest evolution of any of our fitted
parameters. In the case of our two-component fits, we observe a marginal decrease in
the observed number density of the low-mass component over the last $12
\,h_{70}^{-1}\,{\rm Gyr}$, followed by a downturn in the evolution at the highest
redshifts. In contrast to this observed stability, the high mass component evolves
significantly and with more rapidity. The result of this is that the high mass
component begins with little contribution to the mass density, and then builds up to
become the dominant component (in mass) at $\sim\! 8.5\,h_{70}^{-1}\,{\rm Gyr}$
lookback time. This evolution is seen strongly both in our regressed fits and the
individual parameter estimates, with the exception of two outlier bins at $\sim\!
4.5$ and $5.0 \,h_{70}^{-1}\,{\rm Gyr}$, which are circled red in Figure \ref{fig:
results}. These bins are both clipped prior to fitting the regression fit to the
high-mass component evolution of $\phi^\star$, and the former bin is also clipped
prior to fitting the regression fit to the high-mass component evolution of the
stellar mass density.  We note also that this evolution is particularly well matched
to the model presented in \ccite{Leja2015}, although our estimates of other
parameters differ somewhat and therefore our resulting mass density evolution is
somewhat different to that which is presented there. 

{\bbf Further, comparing the goodness-of-fit of both our single and
two-component fits, we see that all bins have reduced-$\chi^2$ values that
overlap. While one may be inclined to argue that this indicates that our fits
are agnostic to the choice of single- or two-component fits in all bins (and,
indeed, this is likely true in most of our higher-redshift bins), it is
important to note the considerable covariance between the two sets of
$\chi_{\rm r}^2$ values. In most bins where we have at least two complementary
datasets, the dominating source of scatter in our presented $\chi_{\rm r}^2$
values is the tension between each of the individual datasets, induced by our
cosmic variance perturbation.  This can be seen in Figure \ref{fig: data bins} and 
in Tables  \ref{tab: 1comp} and \ref{tab: 2comp},
whereby we see jointly higher absolute values and scatter of $\chi_{\rm r}^2$
in bins which are initially in tension; compare, for example, the scatter on
our $\chi_{\rm r}^2$ values in bins $2$ and $3$, or $6$ and $7$, for both sets
of fits. This coherent scatter induces covariance between the $\chi_{\rm r}^2$ 
values in each model, making simple inference regarding model superiority difficult. 
As such, rather than propose a particular model as being better fitting than another, 
we instead focus on what we can learn from the evolution of the two models independently. }  

Finally, we calculate the value of the stellar mass density parameter $\rho_\star$
for each of our fits, {again using the analytic integral of the fits over all masses}. 
The density parameter proves extremely robust to our different
models and fits; all of our fits show a similar evolution and reasonable agreement
with previous work from the literature. This is not surprising as the main
contribution to mass density at each epoch tends to be from $M\eqsim M^\star$
galaxies (and this region is typically well modelled in all the fits). Nonetheless,
the stochasticity is removed in our regressed models, and we can see that these
values follow the literature well. Interestingly we note that our fits find a surplus
of mass density at low redshift with respect to that reported in \ccite{Driver2017},
and in doing so our fits remove any disparity seen between their low redshift bins
and the literature. 

All fit results and regression functions are provided in Appendix \ref{ap: results}. 

\section{Discussion} \label{sec: discussion} 
\begin{figure*}
  \centering
  \includegraphics[width=\textwidth*8/10]{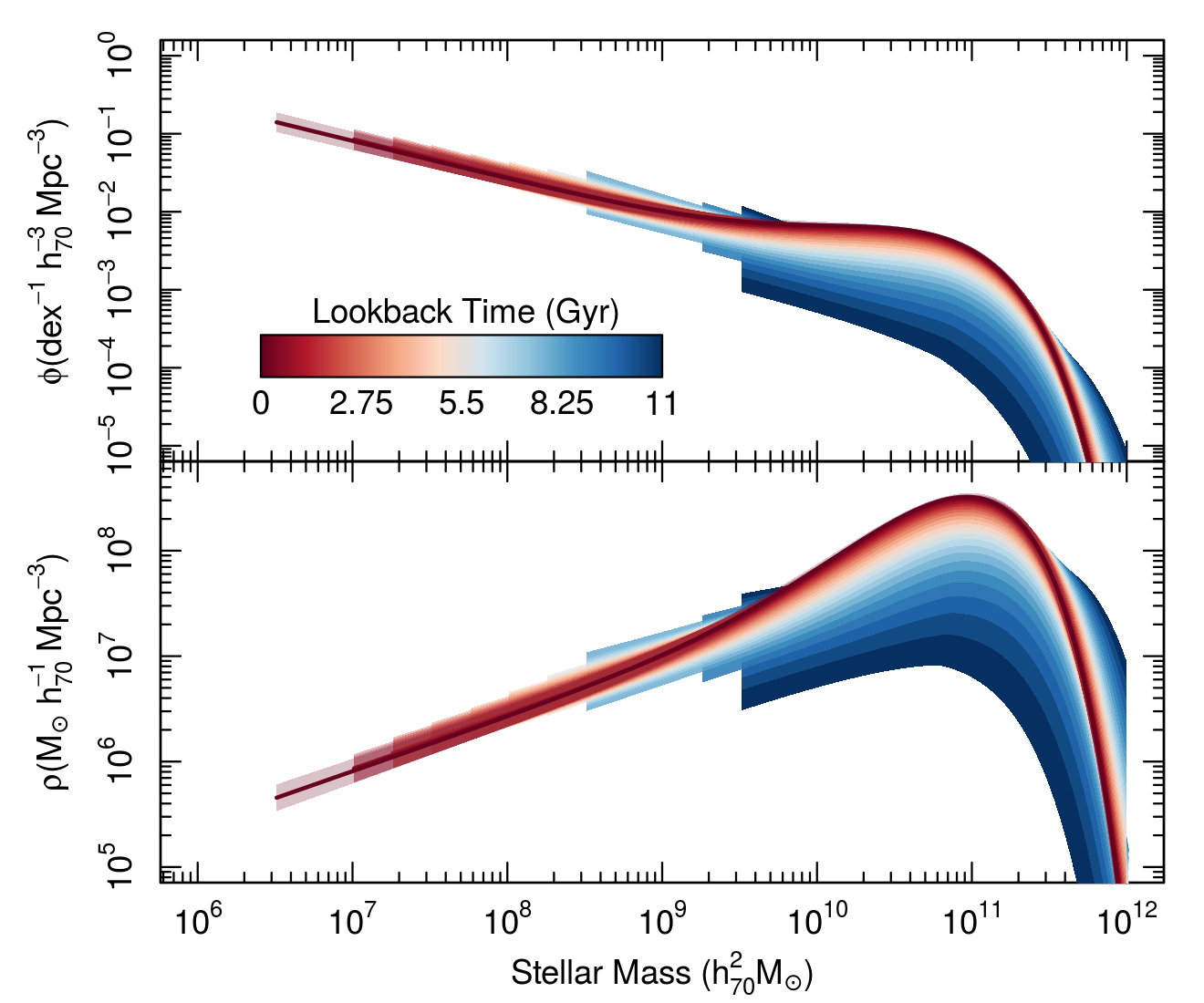}
  \caption{Evolution of the Schechter function over $\sim\! 11\,h_{70}^{-1}\,{\rm
  Gyr}$ (an animated version is provided as supplementary data) Moving forward
  through time, as determined using the regressed Schechter parameters from
  Figure \protect \ref{fig: results}. At each redshift bin we show the model 
  over only the region where we have data. The upper panel shows the evolution of
  galaxy number density $\phi$, while the lower panel shows the evolution of
  the galaxy mass density $\rho$. Uncertainty regions (shaded) show the full
  allowed region of the fit parameters assuming no covariance, and so are
  particularly conservative.  The figure demonstrates clearly the stability of
  the Schechter function faint-end slope over time, with only a modest flatting
  of the number density slope at late times.  The evolution is well described
  by a simple build up of mass in the high-mass component over cosmic time.
  Similarly, the evolution of mass density is seen to be almost  entirely
  driven by the build-up of the high-mass component, around $M^\star$. 
  }\label{fig: evolution}
\end{figure*}
{ In Figure \ref{fig: evolution} we show graphically the evolution of our
fitted Schechter functions, in both number and mass density, over the last 11
Gyr (i.e. in the regime where the evolutional regression fits from Figure
\ref{fig: results} are well constrained). In the figure, we can see the
striking stability of the mass function over this lookback baseline, with only
the high-mass component showing a gradual growth over time. To show this
clearer, we provide an animation of this mass function evolution figure in the
supplementary material. 
}

The stability of the two-component shape parameters, and the observed evolution of
the normalisation parameters, suggests that these two components loosely track two
separate growth mechanisms for galaxies. The low mass component is dominant at early
times after rapidly building up mass in the first few Gyr, but then quickly slows to
a somewhat constant mass density at later epochs. This can be considered to trace
secular evolution of galaxies in the field.  The high-mass component, on the other
hand, demonstrates a lack of mass density at early times but rapidly builds up to
become the dominant component around $t_{\rm trav} \approx 9\,h_{70}^{-1}\,{\rm
Gyr}$. This mode of mass evolution can be considered to trace growth via mergers.
This evolutionary sequence matches well with the mode of mass growth posited in
\ccite{Robotham2014}, where the low mass (disk dominated) end of the GSMF is
populated primarily via secular evolution, and then high mass (bulge dominated)
components grow more significant over time through the galaxy mergers (see their
Figure 17)\footnote{\bbf Note, of course, that this is just one component of the many 
mass growth/loss/redistribution mechanisms in galaxy formation.}. 
We are able to test whether such a growth mechanism matches well with our
observations here by generating a distribution of average growth over the last
$\sim\! 11\,h_{70}^{-1}\,{\rm Gyr}$. Using such a distribution, we will be able to
qualitatively assess, from the shape alone, whether this mechanism is able to explain
the majority of the evolution that we observe. Additionally, we can use the same
distribution to assess whether the mass function evolution that we observe can be
well modelled by a simple steady-state growth, or whether the observed evolution 
exhibits periods of faster/slower/stochastic growth. A constant rate of
growth, for example, may suggest that there exists some regulatory process that
generates a quasi-steady-state relationship between mass growth, destruction, 
and redistribution methods \ccitep[despite observations of higher fractions of
disturbed galaxies at high redshift; see, e.g.,][]{Bridge2010}. 

To test whether the growth we see
is consistent with a constant growth model, we estimate the average fractional growth
of the GSMF per Gyr as:
\begin{equation}\label{eqn: growth}
  \Gamma = \widetilde{\frac{\phi_2-\phi_1}{\phi_1}}\left(t_1-t_2\right)^{-1},
\end{equation}
where $\phi_i$ is the GSMF at lookback time $t_i$, $t_1$ and $t_2$ are chosen as
being the lookback times in two of our GSMF evolution bins, and $\widetilde{x}$
denotes the median of  all bootstrap realisations of $x$ (i.e. this function is
defined using the actual fits in these bins, rather than the regressions).  This
definition has range: 
\begin{equation}\label{eqn: range}
  \Gamma\in\left[-(t_1-t_2)^{-1},\infty\right)\:\forall\,\phi_i\ge0,
\end{equation}
which correspond to the limits where $\phi_1\gg\phi_2$ and $\phi_2\gg\phi_1$
respectively.  This range makes intuitive sense, given the domain
$\phi_i\in\left[0,\infty\right)$; $\phi_2$ can grow to be infinitely larger than
$\phi_1$, but can only lose as much as $\Delta\phi_2 = \phi_1$. 
\begin{figure}
  \centering
  \includegraphics[width=\columnwidth]{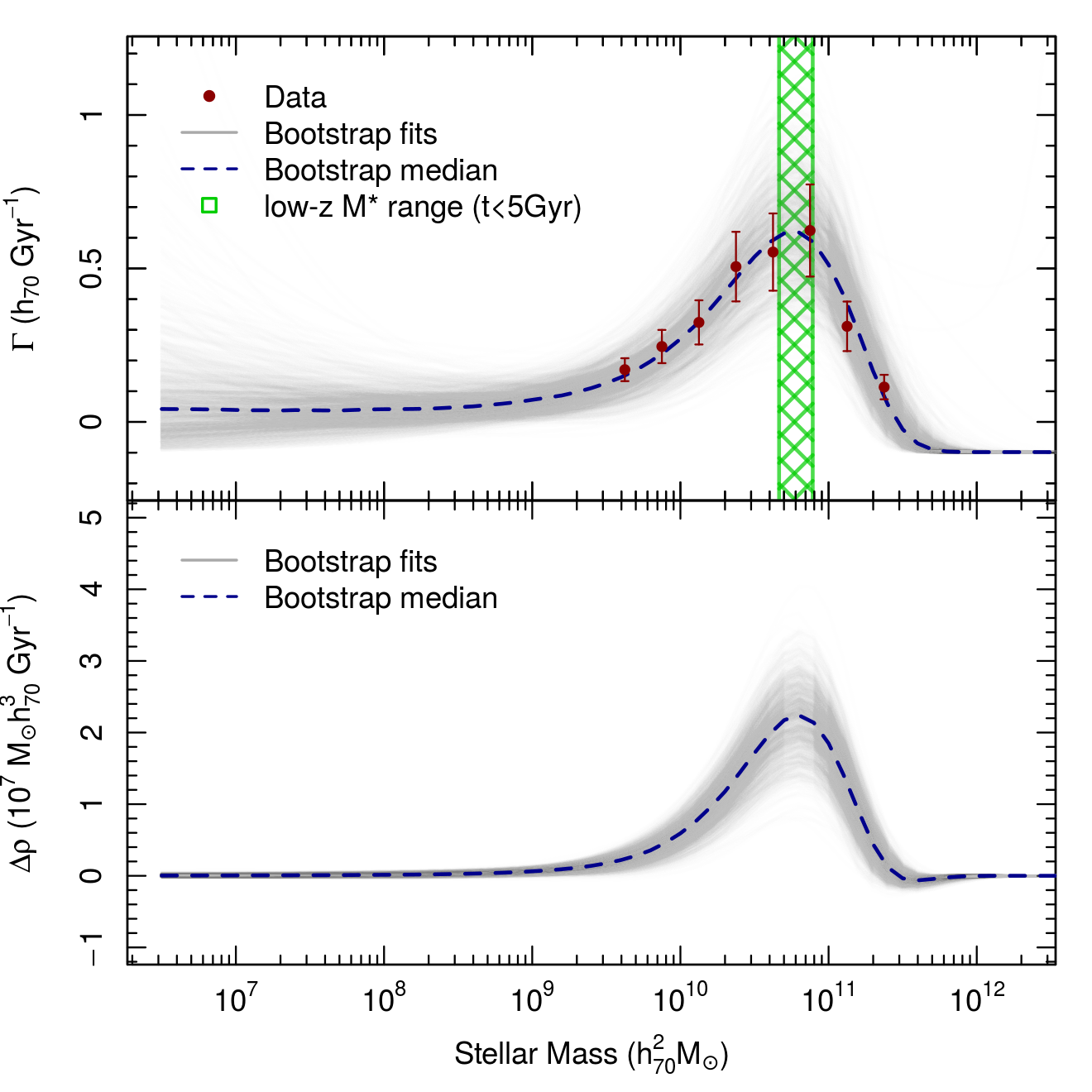}
  \caption{The GSMF average growth function across lookback times
  $t\in\left[0.2,11.0\right]$, shown in fractional growth ($\Gamma$; top) and 
  in mass density ($\rho$; bottom). The distribution, which is calculated as in Equation
  \protect \ref{eqn: growth}, demonstrates the average expected growth of the
  Schechter function number density assuming a uniform growth rate over the entire
  lookback window. The grey lines show the distribution as calculated from the ratio
  of each of the 1001 bootstrap realisations shown in bins $1$ and $15$ of Figure
  \protect \ref{fig: data bins}, with significant transparency.  The red points show
  the ratio of the data in these bins, for the mutually spanned mass range.  The blue
  dashed line is the median of all the bootstrap realisations. The distribution shows
  that, across this range of lookback times, the average growth per Gyr as been
  restricted almost entirely to the high-mass end of the Schechter function; around
  $M^\star$. The low-mass end of the Schechter function shows no significant growth
  (or loss).  At the high-mass end the function converges to the limiting value of
  $-(t_1-t_2)^{-1}\approx-0.1$. 
  }\label{fig: growth}
\end{figure}
We opt to use bins $1$ ($t_2\approx0.61 h_{70}^{-1}{\rm Gyr}$) and $15$ 
($t_1 \approx 10.8 h_{70}^{-1}{\rm Gyr}$) to define 
the growth function, as they span the widest range of lookback time where the
low-mass component is not rapidly evolving in normalisation (see Figure 
\ref{fig: results}). {\bbf Importantly, however, we note that this definition
therefore requires significant extrapolation of the bin $15$ Schechter fits,
well below the lower mass limit of the data in this bin. }

Our average growth function, returned from the individual bootstrap fits to the data, 
is shown in the upper panel of 
Figure \ref{fig: growth}, and the average growth of $\rho_\star$ is shown in the 
lower panel. In both panels we can see a summary of our main conclusion about the
evolution of the GSMF; it shows that there is essentially no change in either the
mass or number density of the low-mass end of the GSMF over the last 
$11\,h^{-1}_{70}{\rm Gyr}$ while the high-mass end the GSMF exhibits strong growth
which peaks at $\Gamma\approx 0.65\,h_{70}^{-1}{\rm Gyr}$, and 
$\Delta\rho_\star \approx 2\times10^7 M_\odot h^3_{70}{\rm Gyr}^{-1}$, 
and is centered on $M^\star$.  
At the highest masses the growth function $\Gamma$ converges on its asymptote
value, as the exponential tail of the low-z mass function beats its compatriot
to 0. 

\begin{figure*}
  \centering
  \includegraphics[width=\textwidth]{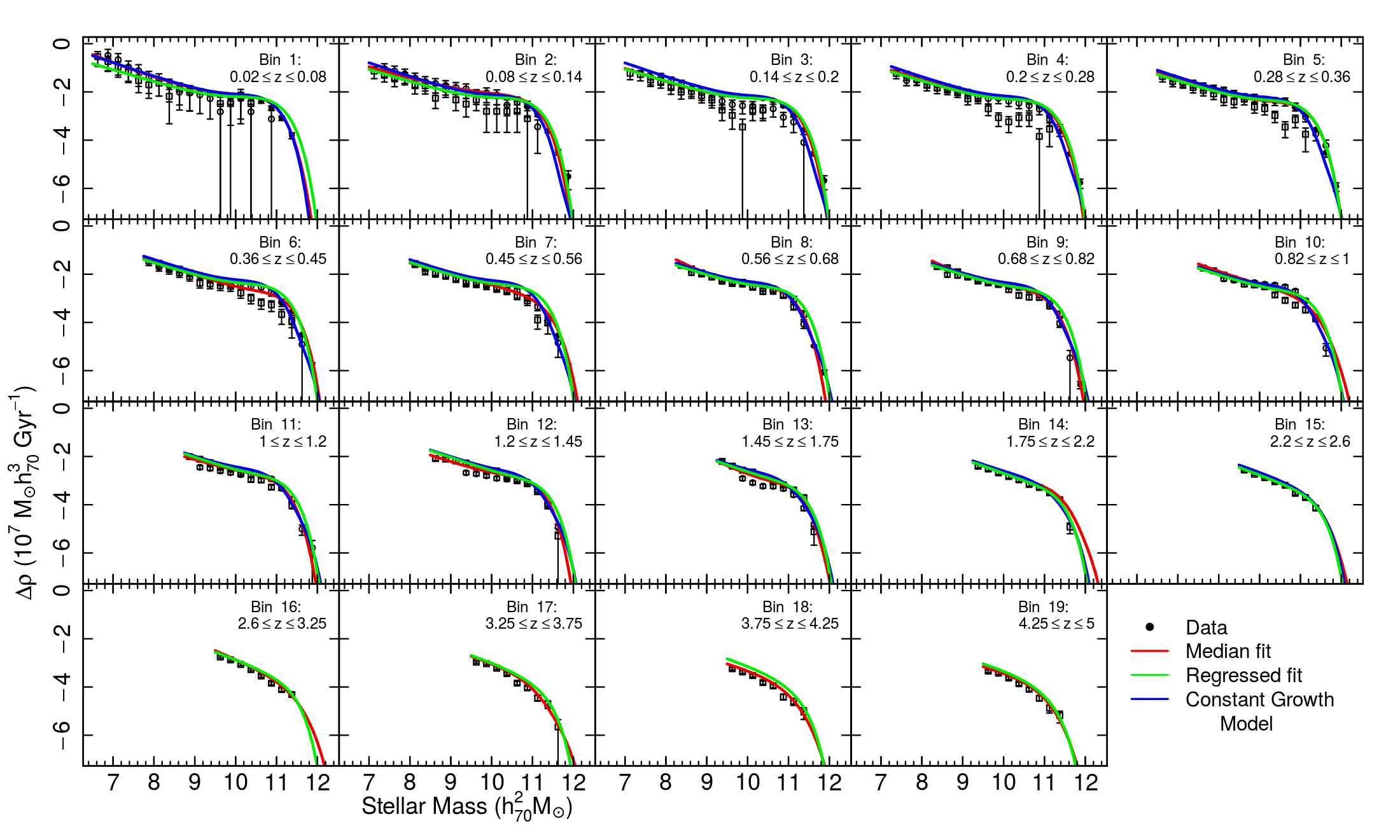}
  \caption{Difference between the three main model types discussed in this work. 
  In each bin, the data are shown in black (reproduced from Figure \protect \ref{fig:
  data bins}), the weighted 
  median of all bootstrap realisations is shown in red, the model returned from our
  best fit regressions (see Figure \ref{fig: results}) is shown in green, and the
  model returned by our constant growth model is shown in blue (not shown in bins
  beyond where the model was defined). At all epochs these
  fits are essentially consistent.
  }\label{fig: models}
\end{figure*}

{ 

The observed stellar mass growth function essentially describes the integrated effect of 
all galaxy stellar mass growth/loss/redistribution 
mechanisms over the $11\,h^{-1}_{70}{\rm Gyr}$s spanned 
by the growth function definition. This would include, but is of course not limited too: 
\begin{itemize}
\item growth due to star-formation from all sources (e.g. secular, merger-driven, etc),
and how the star-formation rate varies over time; 
\item mass lost due to stellar evolution, and how this stellar evolution 
changes with stellar population evolution; and 
\item mass lost to the intragroup/intracluster/intergalactic medium due to stripping/merger events. 
\end{itemize}
Modelling this complex evolution of mass in the universe is a significant task, and 
would require comprehensive modelling of (at least) each of the items listed above. 
Rather than attempt to undertake this task, we instead opt to present our observed mass 
redistribution function as an additional observable that may be of interest to the community, 
and leave this comprehensive modelling for the future.

As an observable, our growth function suggests that the assembly of stellar
mass over the last $\sim 11 h_{70}^{-1}{\rm Gyr}$ has involved an interplay
between the various stellar mass growth/destruction/redistribution mechanisms,
such that no net loss in mass density occurs at any point in the mass function.
As such, this growth function may be verifiable/falsifiable using future
simulations and surveys that endeavour to explore the integrated properties of
mass evolution.  To this end, ongoing and upcoming surveys which will allow the
construction of high-fidelity catalogues of group-scale environments will be
invaluable. Surveys such as the Deep Extragalactic VIsible Legacy
Survey\footnote{\url{devilsurvey.org}} \citep[DEVILS;][]{Davies2018} and the
Wide Area VISTA Extragalactic Survey\footnote{\url{wave-survey.org}}
\citep[WAVES; ][]{Driver2016b} will be able to estimate the integrated
redistribution of mass in a wide range of environments, and will be ideal for
this purpose.

This assumes, however, that the true stellar mass growth in the universe varies 
smoothly. Fortunately, we can simply test} whether our observed average growth function is indeed a reasonable
representation of the observed mass function, and mass density, at each epoch. 
We do this simply by using the average growth model to define a simple 
model GSMF at time $t_i$ as: 
\begin{equation}\label{eqn: model} 
  \phi_{{\rm m},i} = \phi_1\left[1+\Gamma\left(t_1-t_i\right)\right].  
\end{equation} 
To demonstrate the surprising amount of similarity that this simple model
demonstrates to both the data and our best-fit regressed Schechter functions, we
reproduce Figure \ref{fig: data bins} in Figure \ref{fig: models}, except now showing
lines for the median bootstrap model, the best-fit regressed model, and the simple
constant-growth model alongside one-another. The figure demonstrates that the three
models are indeed consistent with each-other, differing most significantly in the
overall Schechter normalisation (where the red bootstrap model lines at each epoch
trace variations in large scale structure). 

\begin{figure*}
  \centering
  \includegraphics[width=\textwidth*2/3]{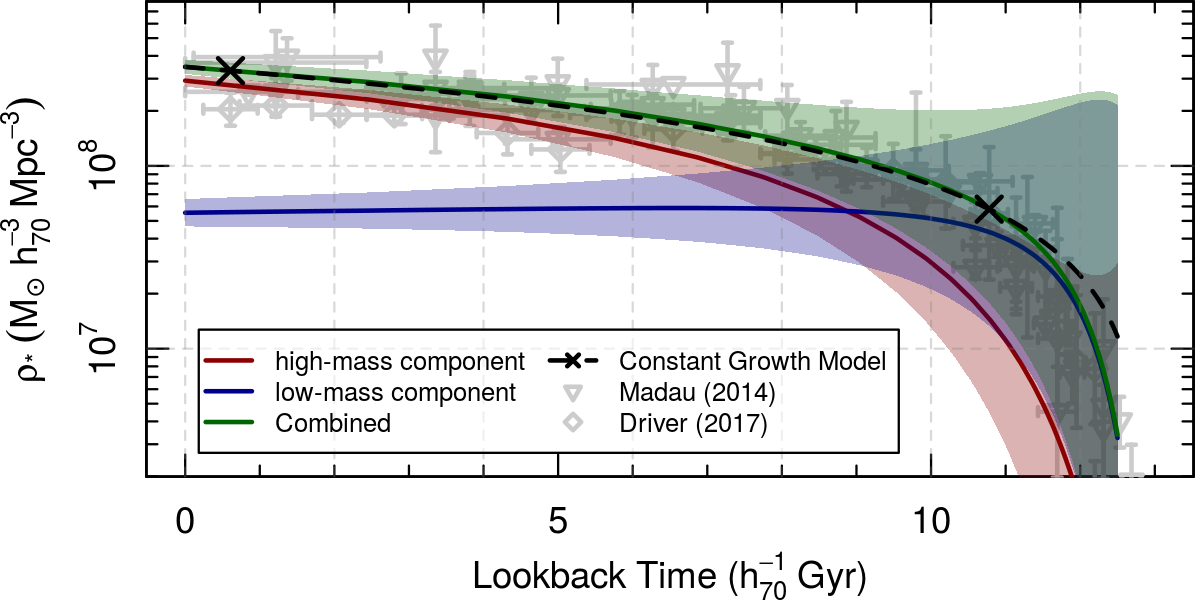}
  \caption{The evolution of the stellar mass density as estimated using our regressed
  fits and with our simple model of constant high-mass growth in the GSMF.  The
  regressed fits (coloured lines) demonstrate the agreement between our data and the
  evolution in the literature (grey points), despite the uncertainties on our fits
  becoming significant at the highest redshifts. Moreover, the simple model of
  constant mass function growth (black dashed line) is also in excellent agreement
  with the literature over the last $\sim\!11\,h_{70}^{-1}\,{\rm Gyr}$. At
  higher-redshift, the simple model is unable to capture the rapid growth of the
  low-mass component, and so the model over-predicts the stellar mass density, as
  expected. Here the growth model $\Gamma^{\rm r}$ has been defined using the
  regressed fits in our bins $1$ and $15$ (shown by the black crosses) so that the
  green and black lines are directly comparable.   
  }\label{fig: mass-density}
\end{figure*}

We then use these simply modelled GSMFs to calculate a model stellar mass
density at every epoch, $\rho^m_\star$, and compare these values to our observed 
evolution of the stellar mass density parameter $\rho_\star$.
These results are shown in Figure \ref{fig: mass-density}, where we reproduce the 
bottom left panel of Figure \ref{fig: results}, except in this instance we show the 
mass density curves derived by our regressed Schechter function parameters, rather
than the observations themselves (as we did in Figure \ref{fig: results}). 
We then overlay the mass density growth returned by our simple constant growth model, 
defined in the same way as above, except that in the figure we
use the regression values at bins $1$ and $15$ to define a growth function
$\Gamma$, rather than the bootstrap fits. This allows us to directly compare
how the constant growth model compares to our regression fits, which make-up our
best-fitting GSMFs at each epoch. 

In the figure, we can see that our regressed
parameters are entirely consistent with the literature, even though the uncertainties
balloon at the highest redshifts\footnote{Recall again, however, that we are assuming
no parameter covariance here. Indeed, comparing the uncertainties between the regressed 
fits to $\rho_\star$ in Figure \ref{fig: results} with those here shows just how much
of an impact the covariance plays in reducing the uncertainties.}
Moreover, we see that the model mass functions $\phi^{\rm r}_{\rm m}$ follow the
evolution of the observed data and of the best-fit regression models essentially
perfectly over the last $\sim\!11\,h_{70}^{-1}\,{\rm Gyr}$. At the highest redshifts,
again, our model does not follow the rapid evolution of the low-mass component and so
over-predicts the mass density somewhat. Nonetheless, this result demonstrates
that with this strikingly simple model of constant growth of the high-mass component
of the GSMF, we are able to reproduce the evolution of the stellar mass density over
the vast majority of the evolution history of the universe and that there has been 
a surprising lack of stocasticity in the overall rate of evolution of the stellar 
mass density. 

\section{Conclusions}\label{sec: conclusions}
In this work we have demonstrated the evolution of \ccite{Schechter1976} function
parameters over $12.5\,h_{70}^{-1}\,{\rm Gyr}$ using the combined sample of GAMA,
G10-COSMOS, and 3D-HST. Using multiple Schechter function fits, we demonstrate that
the single component Schechter function is unlikely to produce reliable fits, even
out to a redshift of 5. Conversely, the two-component Schechter shows impressive
stability of its fitted parameters over the entire redshift range, providing well
constrained parameters at essentially all epochs. We explore the evolution of the
mass function further by regressing the various parameters such that we achieve a
smooth evolution.  Our regressed parameters, in our two-component Schechter fits,
show little to no evolution of the $M^\star$, $\alpha_i$, or low-mass $\phi^\star$
parameters over time, and are especially stable over the last
$\sim\!11.0\,h_{70}^{-1}\,{\rm Gyr}$. Conversely, the high-mass $\phi^\star$
parameter shows strong evolution over the same period. The stability of most
parameters, coupled with the evolution of the high-mass component's normalisation
parameter, suggests a picture of galaxy evolution where these two components broadly
track different mass-evolution mechanisms; the low-mass systems broadly following
secular evolution of galaxies, while high-mass systems are constantly being built up
through merger processes. At the highest redshifts, the low mass component exhibits
somewhat rapid evolution in its normalisation, starting out as the 
mass-dominant component of the GSMF until it is overtaken at
$\sim\!9\,h_{70}^{-1}\,{\rm Gyr}$, when the growing high-mass component becomes the
dominant reservoir of mass. We then test whether the build-up of mass over the last
$11\,h_{70}^{-1}\,{\rm Gyr}$ is well described by a constant rate of mass growth,
finding that this is indeed the case, and that a simple model
of the mass function growth is able to perfectly describe the observed evolution of
the stellar mass density parameter over the majority of the evolution history of the
universe. Nonetheless, we {recognise that this mass growth function encodes a highly 
complex array of mass growth/loss/redistribution mechanisms, and that alone may be only 
used as a guiding observable in future complex mass-assembly studies.}  We 
conclude that upcoming deep and highly complete surveys of group-scale environments, at
intermediate to high redshift, will be required in order to determine the mechanisms
driving the observed growth of stellar mass.      
\section{Acknowledgements} 
 {We thank the anonymous referee for their thorough reading of our work, and for 
 their many helpful comments. }  
 GAMA is a joint European-Australasian project based around a spectroscopic campaign
 using the Anglo-Australian Telescope. The GAMA input catalogue is based on data
 taken from the Sloan Digital Sky Survey and the UKIRT Infrared Deep Sky Survey.
 Complementary imaging of the GAMA regions is being obtained by a number of
 independent survey programmes including GALEX MIS, VST KiDS, VISTA VIKING, WISE,
 Herschel-ATLAS, GMRT and ASKAP providing UV to radio coverage. GAMA is funded by the
 STFC (UK), the ARC (Australia), the AAO and the participating institutions. The GAMA
 website is \url{http://www.gama-survey.org/}. Based on observations made with ESO
 Telescopes at the La Silla Paranal Observatory under programme ID 179.A-2004.
 This work uses photometric data measured using {\sc lambdar} \citep{LAMBDAR}, and higher order data
 products from SEDs measured using {\sc magphys} \citep{MAGPHYS}. Figures were made with the help of
 the {\tt magicaxis} and {\tt celestial} packages \citep{magicaxis,celestial}
 packages in {\sc r} \citep{R}. 
\bibliographystyle{mnras}
\bibliography{library}
\appendix
\section{Fit results \& regression functions}\label{ap: results} 
\vspace{-2in}
\begin{table*}
  \caption{Best fit single-component Schechter function parameters, {and $\chi_{\rm r}^2$ values,} in each redshift bin. {Uncertainties show the asymmetric 1$\sigma$ quantiles on each parameter.} }\label{tab: 1comp}
  \begin{tabular}{cccccccc}
    Bin & $M^\star$  &$\alpha_1$  &$\phi^\star_1$&$\rho_\star$ & $\chi_{\rm r}^2$ \\
    \hline
      1 & $10.86^{+0.056}_{-0.039} $   & $-1.134^{+0.033}_{-0.067}$ &  $-2.491^{+0.1}_{-0.128}  $ &  $8.419^{+0.083}_{-0.089}$ & $1.129_{-0.638}^{+0.972}$\\
      2 & $10.934^{+0.056}_{-0.063}$   & $-1.172^{+0.14}_{-0.115} $ &  $-2.471^{+0.118}_{-0.146}$ &  $8.51^{+0.054}_{-0.059} $ & $1.089_{-0.416}^{+0.612}$\\
      3 & $10.918^{+0.052}_{-0.06} $   & $-1.066^{+0.209}_{-0.127}$ &  $-2.471^{+0.096}_{-0.127}$ &  $8.449^{+0.053}_{-0.055}$ & $2.654_{-0.640}^{+1.110}$\\
      4 & $10.947^{+0.029}_{-0.042}$   & $-1.133^{+0.148}_{-0.096}$ &  $-2.538^{+0.067}_{-0.064}$ &  $8.443^{+0.037}_{-0.04} $ & $2.461_{-0.668}^{+0.944}$\\
      5 & $10.994^{+0.028}_{-0.044}$   & $-1.131^{+0.195}_{-0.106}$ &  $-2.614^{+0.067}_{-0.064}$ &  $8.409^{+0.043}_{-0.049}$ & $2.739_{-0.805}^{+1.418}$\\
      6 & $11.101^{+0.022}_{-0.028}$   & $-1.282^{+0.124}_{-0.065}$ &  $-2.953^{+0.048}_{-0.047}$ &  $8.247^{+0.037}_{-0.053}$ & $2.743_{-0.772}^{+1.379}$\\
      7 & $11.139^{+0.033}_{-0.028}$   & $-1.387^{+0.025}_{-0.025}$ &  $-3.181^{+0.094}_{-0.091}$ &  $8.129^{+0.061}_{-0.071}$ & $0.775_{-0.319}^{+0.790}$\\
      8 & $10.956^{+0.044}_{-0.033}$   & $-1.35^{+0.03}_{-0.042}  $ &  $-2.967^{+0.119}_{-0.169}$ &  $8.125^{+0.076}_{-0.101}$ & $2.185_{-0.664}^{+0.823}$\\
      9 & $10.985^{+0.048}_{-0.019}$   & $-1.372^{+0.018}_{-0.023}$ &  $-3.059^{+0.082}_{-0.098}$ &  $8.097^{+0.062}_{-0.075}$ & $2.008_{-0.649}^{+0.837}$\\
     10 & $11.212^{+0.073}_{-0.058}$   & $-1.486^{+0.024}_{-0.029}$ &  $-3.334^{+0.097}_{-0.105}$ &  $8.124^{+0.061}_{-0.069}$ & $3.523_{-1.135}^{+1.653}$\\
     11 & $11.082^{+0.048}_{-0.039}$   & $-1.319^{+0.024}_{-0.023}$ &  $-3.203^{+0.075}_{-0.081}$ &  $8.006^{+0.057}_{-0.065}$ & $2.630_{-0.763}^{+1.078}$\\
     12 & $11.081^{+0.06}_{-0.035} $   & $-1.297^{+0.031}_{-0.035}$ &  $-3.256^{+0.08}_{-0.083} $ &  $7.947^{+0.056}_{-0.063}$ & $2.987_{-1.209}^{+2.338}$\\
     13 & $11.167^{+0.062}_{-0.055}$   & $-1.398^{+0.069}_{-0.077}$ &  $-3.498^{+0.097}_{-0.134}$ &  $7.844^{+0.052}_{-0.055}$ & $6.571_{-2.919}^{+3.913}$\\
     14 & $11.495^{+0.005}_{-0.131}$   & $-1.528^{+0.023}_{-0.018}$ &  $-3.74^{+0.091}_{-0.092} $ &  $7.983^{+0.077}_{-0.077}$ & $0.271_{-0.150}^{+0.286}$\\
     15 & $11.498^{+0.002}_{-0.234}$   & $-1.664^{+0.094}_{-0.037}$ &  $-4.022^{+0.331}_{-0.07} $ &  $7.896^{+0.035}_{-0.044}$ & $1.235_{-0.646}^{+0.969}$\\
     16 & $11.471^{+0.029}_{-0.256}$   & $-1.726^{+0.022}_{-0.012}$ &  $-4.329^{+0.181}_{-0.069}$ &  $7.613^{+0.053}_{-0.048}$ & $4.006_{-1.867}^{+2.749}$\\
     17 & $11.5^{+0.00}_{-0.011}   $   & $-1.78^{+0.00}_{-0.009}  $ &  $-4.623^{+0.064}_{-0.104}$ &  $7.459^{+0.099}_{-0.075}$ & $5.835_{-2.561}^{+2.818}$\\
     18 & $11.122^{+0.378}_{-0.503}$   & $-1.648^{+0.272}_{-0.133}$ &  $-4.453^{+0.517}_{-0.46} $ &  $7.051^{+0.164}_{-0.159}$ & $2.547_{-1.331}^{+1.692}$\\
     19 & $11.116^{+0.027}_{-0.012}$   & $-1.634^{+0.029}_{-0.042}$ &  $-4.496^{+0.036}_{-0.08} $ &  $7.006^{+0.047}_{-0.052}$ & $1.861_{-0.888}^{+1.555}$\\
    \hline
  \end{tabular}
\end{table*}

\begin{table*}
  \caption{Best fit 2-component Schechter function parameters, {and median $\chi_{\rm r}^2$ values,} in each redshift bin. {Uncertainties show the asymmetric 1$\sigma$ quantiles on each parameter.} }\label{tab: 2comp}
  \begin{tabular}{cccccccc}
    Bin & $M^\star$  &$\alpha_1$  &$\alpha_2$  &$\phi^\star_1$&$\phi^\star_2$&$\rho_\star$ & $\chi_{\rm r}^2$ \\
    \hline
      1 & $10.68^{+0.079}_{-0.077} $ & $-0.515^{+0.206}_{-0.319}$ &  $-1.517^{+0.179}_{-0.204}$ & $-2.304^{+0.117}_{-0.12} $ & $-3.13^{+0.335}_{-0.506} $ &  $8.455^{+0.082}_{-0.093}$ & $0.531_{-0.275}^{+0.512}$ \\
      2 & $10.8^{+0.044}_{-0.043}  $ & $-0.612^{+0.133}_{-0.288}$ &  $-1.457^{+0.101}_{-0.11} $ & $-2.377^{+0.093}_{-0.103}$ & $-3.037^{+0.205}_{-0.711}$ &  $8.526^{+0.05}_{-0.065} $ & $0.836_{-0.329}^{+0.728}$ \\
      3 & $10.819^{+0.033}_{-0.028}$ & $-0.646^{+0.103}_{-0.142}$ &  $-1.507^{+0.049}_{-0.117}$ & $-2.39^{+0.06}_{-0.071}  $ & $-3.452^{+0.191}_{-0.373}$ &  $8.457^{+0.046}_{-0.049}$ & $1.982_{-0.641}^{+0.847}$ \\
      4 & $10.837^{+0.019}_{-0.014}$ & $-0.645^{+0.017}_{-0.082}$ &  $-1.516^{+0.013}_{-0.104}$ & $-2.456^{+0.047}_{-0.047}$ & $-3.384^{+0.169}_{-0.357}$ &  $8.431^{+0.034}_{-0.038}$ & $1.780_{-0.534}^{+0.754}$ \\
      5 & $10.88^{+0.013}_{-0.026} $ & $-0.55^{+0.117}_{-0.043} $ &  $-1.536^{+0.035}_{-0.054}$ & $-2.555^{+0.038}_{-0.041}$ & $-3.4^{+0.125}_{-0.184}  $ &  $8.388^{+0.038}_{-0.038}$ & $1.592_{-0.558}^{+1.207}$ \\
      6 & $10.967^{+0.014}_{-0.017}$ & $-0.487^{+0.052}_{-0.01} $ &  $-1.586^{+0.095}_{-0.019}$ & $-2.936^{+0.048}_{-0.072}$ & $-3.541^{+0.268}_{-0.17} $ &  $8.192^{+0.05}_{-0.047} $ & $1.561_{-0.574}^{+1.294}$ \\
      7 & $11.065^{+0.079}_{-0.067}$ & $-0.487^{+0.055}_{-0.046}$ &  $-1.423^{+0.025}_{-0.118}$ & $-3.632^{+0.409}_{-{\rm Inf}}  $ & $-3.241^{+0.122}_{-0.247}$ &  $8.115^{+0.078}_{-0.079}$ & $0.904_{-0.385}^{+0.705}$ \\
      8 & $10.805^{+0.039}_{-0.021}$ & $-0.782^{+0.136}_{-0.133}$ &  $-1.98^{+0.413}_{-0.02}  $ & $-2.659^{+0.104}_{-0.148}$ & $-4.205^{+0.922}_{-0.127}$ &  $8.508^{+4.728}_{-0.299}$ & $2.561_{-0.661}^{+0.810}$ \\
      9 & $10.881^{+0.04}_{-0.087} $ & $-0.858^{+0.062}_{-0.154}$ &  $-1.838^{+0.254}_{-0.126}$ & $-2.84^{+0.136}_{-0.172} $ & $-3.977^{+0.507}_{-0.356}$ &  $8.204^{+0.152}_{-0.095}$ & $2.495_{-0.792}^{+1.078}$ \\
     10 & $11.06^{+0.191}_{-0.126} $ & $-0.986^{+0.1}_{-0.114}  $ &  $-1.591^{+0.091}_{-0.209}$ & $-3.321^{+0.332}_{-{\rm Inf}}  $ & $-3.532^{+0.186}_{-0.291}$ &  $8.138^{+0.064}_{-0.062}$ & $3.799_{-1.175}^{+1.804}$ \\
     11 & $10.862^{+0.064}_{-0.074}$ & $-0.522^{+0.177}_{-0.155}$ &  $-1.488^{+0.043}_{-0.05} $ & $-3.07^{+0.102}_{-0.092} $ & $-3.409^{+0.092}_{-0.121}$ &  $8.025^{+0.053}_{-0.066}$ & $2.564_{-0.700}^{+1.184}$ \\
     12 & $10.857^{+0.039}_{-0.052}$ & $-0.525^{+0.102}_{-0.097}$ &  $-1.525^{+0.038}_{-0.111}$ & $-3.049^{+0.127}_{-0.09} $ & $-3.589^{+0.105}_{-0.151}$ &  $7.974^{+0.058}_{-0.062}$ & $2.361_{-0.954}^{+1.984}$ \\
     13 & $10.896^{+0.134}_{-0.131}$ & $-0.36^{+0.511}_{-0.552} $ &  $-1.653^{+0.093}_{-0.347}$ & $-3.295^{+0.185}_{-0.145}$ & $-3.785^{+0.244}_{-0.194}$ &  $7.891^{+0.109}_{-0.063}$ & $6.656_{-3.263}^{+4.003}$ \\
     14 & $11.041^{+0.248}_{-0.075}$ & $-0.351^{+0.041}_{-0.027}$ &  $-1.589^{+0.045}_{-0.073}$ & $-3.478^{+0.158}_{-0.499}$ & $-3.648^{+0.099}_{-0.107}$ &  $7.98^{+0.072}_{-0.075} $ & $0.342_{-0.185}^{+0.301}$ \\
     15 & $11.075^{+0.098}_{-0.098}$ & $-0.336^{+0.064}_{-0.074}$ &  $-1.58^{+0.033}_{-0.048} $ & $-4.039^{+0.277}_{-0.389}$ & $-3.62^{+0.062}_{-0.094} $ &  $7.846^{+0.045}_{-0.058}$ & $1.008_{-0.573}^{+1.014}$ \\
     16 & $11.306^{+0.085}_{-0.126}$ & $-0.734^{+0.202}_{-0.138}$ &  $-1.731^{+0.061}_{-0.048}$ & $       -                $ & $-4.169^{+0.185}_{-0.145}$ &  $7.667^{+0.06}_{-0.065} $ & $2.033_{-1.013}^{+1.412}$ \\
     17 & $10.884^{+0.426}_{-0.2}  $ & $-0.154^{+0.457}_{-0.656}$ &  $-1.557^{+0.185}_{-0.199}$ & $       -                $ & $-3.862^{+0.301}_{-0.549}$ &  $7.358^{+0.127}_{-0.078}$ & $4.491_{-2.149}^{+2.901}$ \\
     18 & $10.777^{+0.19}_{-0.167} $ & $0.056^{+0.312}_{-0.365} $ &  $-1.484^{+0.119}_{-0.136}$ & $-4.787^{+0.29}_{-{\rm Inf}}   $ & $-4.005^{+0.197}_{-0.268}$ &  $7.048^{+0.08}_{-0.087} $ & $2.782_{-1.652}^{+2.735}$ \\
     19 & $10.71^{+0.156}_{-0.18}  $ & $0.168^{+0.295}_{-0.277} $ &  $-1.437^{+0.14}_{-0.12}  $ & $-4.846^{+0.331}_{-{\rm Inf}}  $ & $-4.02^{+0.209}_{-0.221} $ &  $6.918^{+0.078}_{-0.069}$ & $3.038_{-1.730}^{+3.205}$ \\
    \hline
  \end{tabular}
\end{table*}

\begin{table*}
  \begin{tabular}{cccccc}
    Fit Type & Parmeter & $A_0$ & $A_1$ & $A_2$  \\
    \hline
    \hline
           & $M^\star$     & $10.791\pm0.050$  & $ 0.558\pm0.056$  & $-0.102\pm0.013$ \\
    Single & $\alpha$      & $-1.160\pm0.060$  & $-0.274\pm0.067$  & $ 0.028\pm0.015$ \\
           & $\phi^\star$  & $-2.455\pm0.069$  & $-0.883\pm0.103$  & $ 0.093\pm0.022$ \\
           & $\rho_\star$  & $ 8.433\pm0.046$  & $-0.273\pm0.063$  & $-0.005\pm0.014$ \\
    \hline
           & $M^\star$       & $10.831\pm0.037$  & $ 0.153\pm0.096$  & $-0.033\pm0.028$ \\
           & $\alpha_1$      & $-0.579\pm0.063$  & $ 0.048\pm0.115$  & $ 0.022\pm0.039$ \\
  Two Comp.& $\alpha_2$      & $-1.489\pm0.038$  & $-0.087\pm0.053$  & $ 0.016\pm0.014$ \\
           & $\phi^\star_1$  & $-2.312\pm0.032$  & $-0.658\pm0.119$  & $ 0.016\pm0.066$ \\
           & $\phi^\star_2$  & $-3.326\pm0.099$  & $-0.158\pm0.103$  & $-0.002\pm0.024$ \\
           & $\rho_{\star,1}$& $ 8.452\pm0.025$  & $-0.554\pm0.091$  & $-0.007\pm0.050$ \\
           & $\rho_{\star,2}$& $ 7.678\pm0.079$  & $ 0.088\pm0.108$  & $-0.050\pm0.026$ \\
           & $\rho_{\star,c}$& $ 8.449\pm0.039$  & $-0.271\pm0.062$  & $-0.012\pm0.015$ \\
    \hline
  \end{tabular}
  \caption{Regression functions displayed in Figure \ref{fig: results} for both single and two-component fits. Fits are quadratic in redshift, where the 
  $A_i$ coefficient applies to the $i^{\rm th}$ power of $z$. }\label{tab: regression}
\end{table*}

\end{document}